\documentclass[preprint2, 10pt]{aastex}

\usepackage{epstopdf}
\usepackage{float}
\usepackage{subfigure}
\usepackage{lineno}
\usepackage{mathtools}
\usepackage{array}

\usepackage{graphicx, subfigure}
\usepackage{float}
\graphicspath{{plots/} }

\received{June 2015}
\revised{revision date}
\accepted{acceptance date}

\shorttitle{Dynamics Resonances}

\shortauthors{Arefieff et al.}

\begin{document}
\title{Dynamics Resonances in Atomic States of Astrophysical Relevance}

\author{K.N.Arefieff}
\affil{St. Petersburg State University, St. Petersburg, 198904 Russiaa}


\author{K. Miculis}
\affil{University of Latvia, Institute of Atomic Physics and Spectroscopy, LV-1586 Riga, Latvia.}

\author{N.N.Bezuglov}
\affil{St. Petersburg State University, St. Petersburg, 198904 Russia and
University ITMO, St. Petersburg, 197101, Russia}

\author{M.S.Dimitrijevi\'c}
\affil{Astronomical Observatory, Volgina 7, 11060 Belgrade 74,
       Serbia and Observatoire de Paris, 92195 Meudon Cedex, France \\and IHIS Techno experts, Batajni\v cki put 23, 11080 Zemun, Serbia}
\email{mdimitrijevic@aob.rs}

\author{A. N. Klyucharev}
\affil{St. Petersburg State University, St. Petersburg, 198904 Russia}
\email{anklucharev@gmail.com}

\author{A.A.Mihajlov, V.A.Sre\'ckovi\'c}
\affil{Institute of physics,Univesity of Belgrade,  P.O. Box 57, 11001, Belgrade, Serbia}

\email{vlada,mihajlov@ipb.ac.rs}

\begin{abstract}

Ionized geocosmic media parameters in a thermal and a subthermal range of energy have a number of unique features. The photoresonance plasma that is formed by optical excitation of the lowest excited (resonance) atomic states is one example of conversion of radiation energy into electrical one. Since spontaneous fluorescence of excited atoms is probabilistic, the description of the radiating quantized system evolution along with photons energy transfer in a cold atoms medium, should include elements of stochastic dynamics. Finally, the chaotic dynamics of a weakly bound Rydberg electron over a grid of the energy levels diagram of a quasi-molecular Rydberg complex provides an excitation migration of the electron forward to the ionization continuum. This work aims at discussing the specific features of the dynamic resonances formalism in the description of processes involving Rydberg states of an excited atom, including features in the fluorescence spectrum partially caused by the quantum defect control due to the presence of statistic electromagnetic fields.
\end{abstract}

\keywords{Rydberg quasi-molecule, stochastic dynamics, spectral lines emission.}

\section{INTRODUCTION}

Publications on astrophysics have traditionally paid attention to the processes involving highly excited (Rydberg) atom: in stellar atmospheres of late spectral types, interstellar nebulae and other space objects, including our solar system. The concentration of periodic table elements in geocosmic entities can exceed their average concentration in the Universe by many orders of magnitude. The examples include the clouds of alkali atoms K and Na in the atmosphere of the Galilean satellite Io of the planet Jupiter and chemi-ionization processes that result in the formation of a wide range of molecular ions. The constants of these processes in the thermal interval of energy can reach values of the order of $10^{-9} \textrm{cm}^3\textrm{s}^{-1}$ (see, for example, \citet{kly07}). Since the characteristics of Rydberg states of atoms and quasi-molecular complexes are with certain exceptions (which are described in terms of a quantum defect theory) close to those of a hydrogen atom, they are nowadays widely used in solving current problems of nonlinear dynamics of quantized systems.

Physics and chemistry of low-temperature plasma are most interested in the processes where a transition of an optical electron into ionization continuum  changes the nature of particles and occurs in thermal collisions. These processes are traditionally considered as an effective channel of molecular ion formation. According to publications, the maximum values of effective constants of the formation of a stable charged complex in binary collisions correspond to the value range of principal quantum numbers $5\le n\le 25$ (see, for instance, \citet{mih12}).

Enrico Fermi made the fundamental contribution into the physics of atomic collisions involving Rydberg atoms (RA). He suggested treating the structure of the Rydberg diatomic quasi-molecular complex as the system of two Coulomb centers and a quasi-free Rydberg electron (RE) on the trajectory orbit \citep{fer34}. Cold ($\sim 90 \mu \textrm{K}$) molecules of alkali atoms were observed in magneto-optical traps.

Today, the study of phenomena that determine evolution and stochastic properties of deterministic Hamiltonian systems is an important branch of nonlinear mechanics with potentially interesting applications in atomic and molecular physics. Research in the field of non-linear mechanics \citep{zas91,sag88} indicates that dynamic chaos conditions should be treated as a typical rather than an exceptional situation for the dynamics of Hamiltonian systems with small number of degrees of freedom. In the physics of atomic collisions under the semi-classical treaty, chaos conditions arise as a result of unstable trajectory orbits of optical electrons. Generally, there are always regions of phase space and parameters of a system of interacting particles, where the Hamiltonian system dynamics is stochastic. The transition from integrable problems of classical dynamics to a system with chaos is accompanied by the emergence of areas in the phase space that are the centers of chaos and form "stochastic layers" and a "stochastic web" \citep{zas91}. Even in a one-dimensional problem with a time-variable external disturbance, the effects of dynamic chaos may occur.

A complete review of data (more than 280 references) obtained before 1995 with regard to the dynamics of the optical electron in an excited hydrogen atom in external microwave and alternating electric fields can be found in \citet{koc95}. Work of \citet{del83} is noteworthy as one of the studies that influenced the development of theoretical and experimental research in the field of dynamic chaos phenomena in  excited atom systems. This research grounded the applicability of quasi-classics in describing the development of dynamic chaos phenomena in a quasi-hydrogen Rydberg alkali atom and showed the difference of the arising process of "stochastic diffusion" from multiphoton and tunnel ionizations in a diatomic quasi-molecular complex. In the diffusion case ,\textit{ }the stochasticity domain in the principal quantum number space is $n_{min}< n< n_{max}$, where $n<n_{min}$ are states with the outer electron regular motion, and $n>n_{max}$ belong to continuous spectrum states defined by Chirikov criterion \citep{chi79}.

Today quasi-classical approximation of quantum mechanics \citep{sag88} involving Bohr-Sommerfeld quantum conditions with its visual interpretation is widely and practically applied in atomic physics. Within its framework, the main results have been obtained so far with regard to the problem of associative ionization occurring in the intermediate state of a Rydberg quasi-molecular diatomic complex \citep{mih12}.

Importantly, the stochastic ionization treaties for quasimolecular collision complexes or under situations when ionization is stimulated by external microwave fields are essentially one-dimensional since they are based on the assumption that the RE orbital momentum  $\vec{L}$ remains invariant in the process of RE diffusion. The issue how $\vec{L}$ evolves in the case of the actual three-dimension quantum systems and whether this evolution can significantly affect the results of one-dimensional approaches is briefly outlined in the present paper. Another important issue is related to the influence of the external statistic electromagnetic fields on the atomic quantum states structure and, as a consequence, on the values of dipole matrix elements. The latter determines the rate constants values in the equations describing radiation/impact kinetics of geocosmic media, so that the quantum defect control by external fields can serve as a means to manipulate both the Rydberg complexes emission spectra and the collision ionization. It is known that direct analogies exist between the processes in the laboratory (Earth's) plasma and geocosmic plasma. In the atmospheres of helium-rich cooling white dwarfs stars, as well as in Earth conditions in the temperature range of $(12-16)\cdot10^3$K with ionization degree of the order of $10^{-3}$, chemi-ionization plays the major role in the balance of ionization-recombination processes with $n\ge10$ \citep{ign08}. Hereinafter the system of atomic units (at.u.) is used.

\section{Dipole resonance ionization model for Rydberg quasi-molecular complexes  (DRM-theory)}

The atoms in the upper excited  states with values of effective quantum numbers $n^{*} \ge 5$ have unique characteristics conditioned by the strong dependence of their parameters on $n^{*}$: polarizability $\sim n^{*7}$, dipole moment $\sim n^{*2}$ etc., making them effective partners in inelastic atomic and molecular collisions.  The dipole resonance ionization mechanism is briefly described below according to \citet{mih12}. So far the absolute majority of calculations of ionization processes occurring in the intermediate state of a Rydberg quasi-molecular complex have been performed within the framework of that ionization model. The Rydberg quasi-molecular complex $A_{2}^{**}$  stabilization or decay with the formation of charged fragments are determined by the set of such parameters as peculiarities of interaction potentials, the nature and the level of particle excitation, and the relative collision energy. The main contribution into the ionization occurs in the range of internuclear distances $R$ at which the weakly bound (outer) electron becomes shared within the parent  complex $A_{2}^{+}$. Charge exchange in that complex induces the generation of an time varying dipole moment $\sim R(t)$ that, in its turn, causes the emergence of a quasi-monochromatic microwave electric field pertubing Rydberg electron (RE) motion in the Keplerian orbit. The corresponding  charge exchange frequency $\omega(R)$ appears to be equal to the exchange interaction $\Delta(R)$ or, equivalently, to the frequency of complete rotations of the low-bound electron orbital motion. The probability of autoionization decay of the complex $A_{2}^{**}$ is proportional to $R^2$ and depends on the value of the RE photoionization cross-section, i.e. on the probability of the RE stimulated transition from the bound state with the binding energy $E_{i}^{*}(n)<0$ into the states of a continuous spectrum with the electron kinetic energy $\varepsilon_{k}=p_{k}^{2}/2 \ge 0$. The emitted electron energy coincides with the energy released during the transition of the initial state of $A_{2}^{**}(R)$ into the state $A_{2}^{+}(R)$, that is reflected in the name "dipole resonance mechanism" of the process.

It should be noted that today the interest in chemi-ionization of Rydberg atoms is experiencing a renaissance due to the need to take them into account in the quantum electronics physics, on the one hand, and on the other hand, due to the negative impact of Rydberg complexes induced by geomagnetic disturbances in the Earth's ionosphere on our biosphere, including the human being per se. Deterministic DRM-theory has played a major role in understanding the processes that underlie ionization mechanisms involving Rydberg atoms. Moreover, the model is valid for both a symmetrical and an asymmetrical process \citep{smi71}. New features of the DRM-theory largely stimulate the development of experimental research, for example, the use of synchrotron radiation to excite molecular clusters \citep{kra10}.

\begin{figure*}
\centerline{\includegraphics[width=\columnwidth, height=0.75\columnwidth]{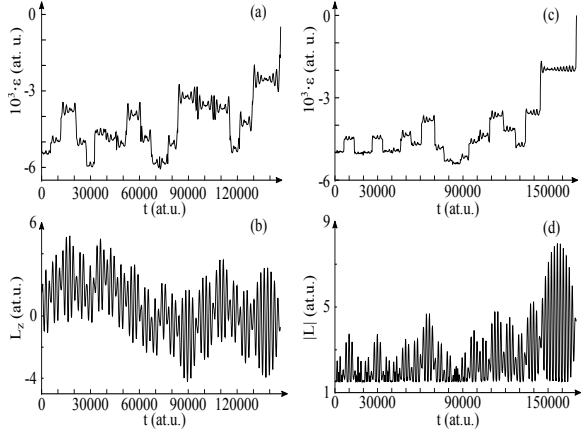}}
\caption{Evolution of the Rydberg electron energy $\varepsilon$ (frames a,c) and its angular momentum $\vec{L}$ (frames b,d) for $10p$ state ($n_0 = 10$, $l=1$) of hydrogen atom. In the case of RE plane motion (frames a, b) frame b exhibits the projection $L_{Z}$ of $\vec{L}$ on the direction $\vec{L}/|\vec{L}|$ orthogonal to the plane of RE motion. Frame d corresponds to three-dimensional RE motion and shoes the length $\vec{L}=|\vec{L}|$ of
the angular momentum.} \label{fig:1}
\end{figure*}

\section{Non-linear dynamic resonances in atomic systems in an external time-depending electric field.}

In the literature, the term "dynamic chaos" in atomic systems refers to the result of the evolution of deterministic (causal) quantized atomic systems, which dynamics over time cannot be accurately predicted. As an example, consider three options:

(i)  Rydberg hydrogen atom or quasi-hydrogen alkali atom in an external electric (electromagnetic) field \citep{del83,kra10}.

(ii)  Rydberg collisional quasi-molecular complex, i.e. Rydberg quasi-molecule.

(iii) The diffusion ionization of Rydberg collision complexes with a strong non-adiabatic connection between electron motion and nuclear motion.

The above cited papers \citet{del83} and \citet{kra10} formulate the traditional approach to the problem of the excited atom behavior in an external microwave electric field, show the applicability of a quasi-classical description for RE dynamics which starts with the initial energy $\varepsilon_{0} $. The field perturbation causes the main change $\Delta n = n - n_0$ in the principal quantum number $n_0$ of the initial state of the atom when the external field angular frequency $\omega_{L}$ coincides with the circular frequency of the electron motion in the Keplerian orbit $\omega_{c}=n^{*-3}$ (resonance). In the case of alkali atoms the effective quantum number $n^{*}$ differs from the principal one $n$ on the value of the quantum defect  $\delta_{l}$: $n^{*}=n-\delta_{l}$. Essentially, the authors of \citet{del83} (see also \citet{gon87}) suggested a stochastic ionization mechanism unparalleled in tunnel ionization or multiphoton ionization, i.e. the diffusion ionization mechanism. In quantum mechanics, this is explained within the model of "multiphoton" resonance, under which the $m$-photon energy coincides with the energy levels separation of an excited particle. Quasi-classical approach connects the stochastization effect with the emergence of a stationary term $U_{k,m}(\varepsilon $) in the perturbing series that formally corresponds to the m-photon resonance when the product $m \omega_L$ coincides with the RE k-overtone: $m \omega_L  =k \omega_c$ where m and k are integer numbers. In this case the value of the excited atom energy $\varepsilon$ begins to oscillate in the vicinity of its "undisturbed" value $\varepsilon_0$ with the width of the nonlinear resonance $\delta \varepsilon$.

\begin{equation}
\label{eq::1}
\delta \varepsilon^2 \sim \omega_{c}U_{k,m}(\varepsilon)/(d \omega_{c}/d\varepsilon),
\quad \varepsilon = \varepsilon_{0}=-1/(2n_{0}^{*2}),
\end{equation}

Two fixed energy levels that have the quantum numbers $n_{0}$ and $n_0 + k$ mix most effectively when m-photon resonance is realized,

\begin{equation}
\label{eq::2}
m \omega_L=\Delta \varepsilon =\varepsilon (n_{0}+k)-\varepsilon(n_0)\approx k \omega_c(n_0),
\end{equation}

\noindent i.e. the m-photon energy $m \omega_L$ coincides with the energy spacing $\approx k \omega_c$ between the levels. In the literature, this situation is called the implementation of dynamic nonlinear resonance for the "resonance" energy $\varepsilon _{k,m}=\varepsilon(n_0)$. Only in this case, the external pertubation can significantly distort the initial (unperturbed) motion of the electron. Note also the root (non-linear) dependence of $\delta \varepsilon$ on the value of the disturbing potential.

The emergence of the finite width of nonlinear resonance for nonlinear systems ($d \omega_c/d \varepsilon \neq 0)$ is crucial for the formulation of global (strong) chaos conditions. Generally speaking, in the energy space with a given external excitation, a discrete set of "resonant" energies $\varepsilon_{k,m}$, which are conditioned by nonlinear resonances of different orders, can be realized (along with the corresponding quantum numbers $n_{k,m}$). In the case of a weak excitation the widths $\delta \varepsilon$ of neighboring resonances do not overlap, i.e. $\delta \varepsilon < \Delta \varepsilon$. If the system motion starts with the initial energy value $\varepsilon = \varepsilon_0$, the system always remains "bound" within the width $\delta \varepsilon$. The situation changes radically when the regions of neighboring nonlinear resonances begin to overlap, i.e.  $\delta \varepsilon > \Delta \varepsilon$. Now the electron can "diffuse" in the neighborhood of any nonlinear resonance and go away from its initial position as far as can be in the energy space. The resonance overlap condition is known in the literature as Chirikov criterion for the onset of the global chaos \citep{chi79}.

The ratio $K= \delta \varepsilon / \Delta \varepsilon$ (Chirikov parameter) equals one was shown to be the threshold of the global chaos realization. Note that the widths of dynamic resonances and, consequently, the effects of stochastic dynamics are directly related to the matrix elements of perturbation operators. This observation is confirmed by the data in \citet{bez14b}, where the coefficients of the light-induced RE diffusion equation in an external microwave field are explicitly expressed using dipole matrix elements of an external field. Some results of the theory of dynamical chaos, which are useful for carrying out relevant assessments may be found in works of \citet{koc95,del83,kra10,kau87,bez01,kau99} and \citet{bez02}.

Studies of the dynamic chaos in quantized systems place high demands on the numerical algorithms stability at large times. Papers of \citet{efi14a,efi14b} and \citet{bal07} suggest to use geometric integration methods of \citet{hai99} in describing evolution of a Rydberg atom in external fields. These methods combine the enhanced stability of symplectic methods of differential equations numerical solution with the efficiency (calculation speed). The improved method of split evolutions was also applied in calculations on the base of Floquet technique \citep{chu04} that allows extending the algorithms used for nonstationary problems. As a result, an algorithm for calculating RE trajectories in the Coulomb potential perturbed with a microwave electrical field was elaborated \citep{efi14a,efi14b}.

Below we show in Fig.~1 numerical results on the chaotic diffusion dynamics of RE in a hydrogen atom which evolves from the initial state 10P (with the principal $n_0=10$, and orbital $l=1$ quantum numbers) to the ionization continuum in the presence of the external microwave field $\vec{E}=\vec{E}_{0}\cos(\omega_{L}t)$. The field frequency is taken as $\omega_L=3/10^3$ at.u. The field amplitude $E_{0} =|\vec{E_{0}}|$ is chosen to exceed the critical value $E_{c} =2/(49n_{0}^{4})$, \citep{del83,kra10} for the onset of the dynamic chaos regime. When the amplitude $E_0 > E_c$, all dynamic resonances occurring in the region $n > n_0$ overlap, so that RE may free-diffuse into continuum. Under the semi-classical approach the description of the wave-function evolution is carried out in terms of the RE classical trajectories characteristics such as the energy $\varepsilon$ and the angular momentum $\vec{L}$. The initial value $L_{0}$  of $L= |\vec{L}|$ is determined, for instance, via the initial orbital quantum number $l$  with the semi-classical formula $L_{0}= l +0.5 = 1.5$ \citep{del83,ree06} while the initial energy $\varepsilon_{0}$  of the trajectory is obtained via Eq.~(1). We are concerned mainly with the situations when $L$ tends to change significantly, i.e. states with various $l$ mix. Data of Fig. 1 correspond to two basic configurations resulting in the maximum possible variations of $L$: (i) RE plane motion ($\vec{L_{0}} \perp  \vec{E_{0}} $, frames a, b) with the amplitude $E_{0} =16.4/(49n_{0}^{4}) $ and (ii) RE three-dimensional motion ($\vec{L_{0}} \parallel  \vec{E_{0}} $, frames c, d) with $E_{0} =12.9/(49n_{0}^{4})$.

However, data in Fig.~1 do not negate the results of the initial one-dimensional diffusion ionization theory of \citet{del83,kra10} that takes into account the chaotic migration of RE only along the energy spectrum without regard to changing \textit{L}. Importantly, it is well seen that \textit{L} variations in the process of diffusion remains below its critical value $L_c \sim l =10$, exceeding of which results in deviations from one-dimensional theory \citep{del83,kra10}. As it turns out with lower values of $L <L_c$, the diffusion rate constants calculated using the equations of Fokker-Planck type practically does not depend on $L$ \citep{kra10}, so that $L$ variations with amplitude less than $L_c$ do not affect the one-dimensional approach for RE stochastic ionization.

\begin{figure}
\centerline{\includegraphics[width=1\columnwidth, height=\columnwidth]{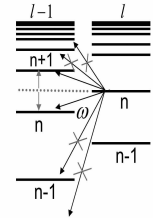}}
\caption{Schematic energy levels diagram for two series \{l -- 1\} and \{ l \} that realize the F\"{o}ster resonance.
The levels structure is identical to that for a
three-dimensional oscillator and in accordance with the oscillator selection rules \citep{som34}
allows dipole optical transitions  between adjacent levels only.} \label{fig:2}
\end{figure}
\begin{figure}
\centerline{\includegraphics[width=\columnwidth, height=0.85\columnwidth]{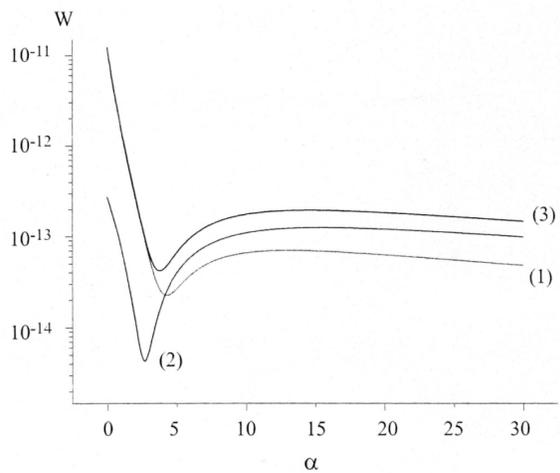}}
\caption{Partial probabilities $W_{p \rightarrow s}$ (curve 1) and $W_{p \rightarrow d}$ (curve 2) (atomic units) of spontaneous transitions for the state 13p (n=13, l=1) as functions on the model parameter $\alpha $ entering into Sommerfeld potential $U_{\alpha}(r)$. Curve 3 shows the behavior of the total spontaneous decay rate $W_{p}(n)= W_{p \rightarrow s}+ W_{p \rightarrow d}$.} \label{fig:3}
\end{figure}
\section{Specific features of stochastic and radiative processes under double Stark resonance.}

Today the research in spectral composition of Rydberg atoms is one of the main trends in physics of Earth's atmosphere and astrophysics. In the literature, the case where one of the levels of \textit{l-}series of an atom is located in energy space exactly between the two levels of adjacent series (\textit{l}~--~1) or (\textit{l}~+~1) is called F\"{o}ster resonance or double Stark resonance \citep{wal05}. This configuration corresponds to a two-photon resonance for a cascade transition $(l-1, n)\rightarrow (l, n) \rightarrow (l-1, n-1)$ (see Fig.~2). Such situation occurring when $\delta_{l-1}-\delta_{l}=0.5$  in terms of the quantum defect $\delta$ is typical for Rydberg alkali atoms (in the case of Na, for instance, $\delta_{l=0}- \delta_{l=1}\approx 0.504$), while it is absent in the hydrogen atom. Today, the literature on spectroscopy considers this phenomenon as a prospective option for manipulating cold atoms in a laser field \citep{sho11,ree06}. The configuration of atomic levels that meets the case of double Stark resonance presented in Fig.~2 is similar to that of a three-dimensional quantum oscillator according to the second Bohr correspondence rule \citep{som34}, where "long" optical transitions are forbidden (see Fig. 3). In other words, it is the matter of most of the transitions being blocked due to a significant decrease in the values of their dipole matrix elements. The both electron diffusion over the grid of quasi-crossing energy curves of a Rydberg complex and radiative processes should be damped under the realization of F\"{o}ster resonance.

As an illustration, we consider RE features in alkali atoms which can be modeled with Sommerfeld potential $U_{\alpha } (r)=-1/r+\alpha /(2r^{2} )$ \citep{som34}, where $\alpha$ is a parameter of our model. In the case of Hydrogen atom the parameter $\alpha$ is zero. For alkali atoms $\alpha $-values should be found from spectroscopic data using the following relation between the quantum defect $\delta_{l}$ of $l$-atomic series and the parameter $\alpha$: $\delta_{l}= l+0.5-[(l+0.5)^2+\alpha]^{1/2}$. The conditions of F\"{o}ster resonance implementation $\delta_{l-1}-\delta_{l} =0.5$ for a Sommerfeld atom corresponds to the parameters $\alpha$ values $\alpha_{l} = 3(l^2-1/16)$ \citep{zak11}. In the particular case of the $\{p, s\}$ series it yields $\alpha_{p} = 2.81$, whereas $\alpha_{d}  = 11.8$ for the $\{d, p\}$ series. The particularities of stochastic ionization were discussed in work of \citet{zak11} where it was demonstrated a threefold increase in the time of the diffusion ionization in the vicinity of the double Stark resonance. Here we present calculations (see Fig.~3) for the total probabilities $W_{l\rightarrow l'}(n)$ of the spontaneous transitions of $nl$-state at all lower lying levels of the fixed \textit{l'}-series (\textit{l'} = \textit{l} $\pm$ 1). We choose the initial Rydberg state of $p$-series (\textit{l} = 1) with \textit{n} = 13. It is well seen a significant decrease (at three orders of magnitude) of spontaneous transitions probabilities in the vicinity of the critical value $\alpha_{p}$ = 2.81, where blocking the dipole matrix elements takes place due to F\"{o}ster resonance (see Fig.~2). Those important facts should be properly accounted for in the radiation-collisional kinetics in the atmospheres of celestial objects under the presence of quasi-static electric/magnetic fields resulting in essential Stark/Zeeman shifts of atomic levels. One of the consequences, for instance, may be dramatic redistribution of Rydberg states $(n \sim 10) $ populations with the subsequent changes in IR emission spectra.

\section{Conclusion}

In the literature, the term "chaotic" refers to quantum-mechanical or quasi-classical ensembles. It implies a probabilistic description of physical phenomena which cannot be reduced to a conventional notation of an individual wave functions or a trajectory. The main reason for the conversion of a regular atomic system into chaotic one under the influence of external time-varying forces is mainly connected with the emergence of the so-called "dynamic nonlinear resonances". Their mutual overlapping results in the emergence of a global dynamical chaos. In this paper, the stochasticity is considered with regard to diffusion ionization, arising due to complicated energy levels of a Rydberg quasi-molecular collisional complex or stimulated by an external weak microwave field. We discussed as well a possibility of blocking dipole optical transitions in a Rydberg atom due to the quantum states shifts under the presence of statistic electromagnetic fields and describe the corresponding dramatic changes in both the dynamic chaos regime and kinetics of radiative processes. Important specific futures of the double Stark resonance are outlined that are of interest for the interpretation of the observed fluorescence spectra in astrophysics.

\acknowledgments

This work was supported by Government of RF, Grant 074-U01 and by EUFP7 Centre of Excellence FOTONIKA-ZV (REGPOT-CT-2011-285912-FOTONIKA). Also, the authors are thankful to the Ministry of Education, Science and Technological Development of the Republic of Serbia for the support of this work within the projects 176002, III44002
\\

\newcommand{\noopsort}[1]{} \newcommand{\printfirst}[2]{#1}
  \newcommand{\singleletter}[1]{#1} \newcommand{\switchargs}[2]{#2#1}

\end{document}